\documentclass{ifacconf}
\usepackage{graphicx}
\usepackage{natbib}
\usepackage[T1]{fontenc}
\usepackage{amssymb,amsfonts}
\usepackage{xcolor}
\usepackage{mathtools}
\usepackage{enumitem}
\usepackage{svg}
\newcommand{\tr}{\operatorname{tr}}
\newcommand*{\tran}{^{\mkern-1.5mu\mathsf{T}}}
\newtheorem{theorem}{Theorem}

\newtheorem{Lemma}{Lemma}
\newtheorem{Definition}{Definition}
\newtheorem{Remark}{Remark}

\newtheorem{Assumption}{Assumption}
\definecolor{crimsonglory}{rgb}{0.75, 0.0, 0.2}

\graphicspath{{figs/}}
\begin{document}
\begin{frontmatter}

\title{Second-Order Policy Gradient Methods for the Linear Quadratic Regulator}

 \author[First]{Amirreza Valaei}
 \author[Second]{Arash Bahari Kordabad}
 \author[Second]{Sadegh Soudjani}

 \address[First]{Aktus AI, San Mateo, California (e-mail: \texttt{amirreza@aktus.ai})}
 \address[Second]{Max Planck Institute for Software Systems (MPI-SWS),
 Kaiserslautern, Germany (e-mails: \texttt{\{arashbk,sadegh\}@mpi-sws.org}).}

  \begin{abstract}
      Policy gradient methods are a powerful family of reinforcement learning algorithms for continuous control that optimize a policy directly. However, standard first-order methods often converge slowly. Second-order methods can accelerate learning by using curvature information, but they are typically expensive to compute. The linear quadratic regulator (LQR) is a practical setting in which key quantities, such as the policy gradient, admit closed-form expressions. In this work, we develop second-order policy gradient algorithms for LQR by deriving explicit formulas for both the approximate and exact Hessians used in Gauss--Newton and Newton methods, respectively. Numerical experiments show a faster convergence rate for the proposed second-order approach over the standard first-order policy gradient baseline.
  \end{abstract}


\end{frontmatter}
%===============================================================================

\section{Introduction}
Optimal control problems play a fundamental role in diverse domains of engineering and science, with applications ranging from robotics and autonomous systems~\citep{dexterousmanipulation,MOHEBIFARD2019} to economic systems~\citep{ApplicationsOptimalControl,Dorfman1969EconomicInterpretation}. The objective is typically to design a feedback policy that minimizes a long-term cost while satisfying the dynamics of the underlying system.

Policy gradient methods optimize parameterized policies via estimated performance gradients and scale to continuous, high-dimensional action spaces~\citep{silver2014deterministic}. Conventional policy gradient methods typically use first-order optimization~\citep{silver2014deterministic,nocedal2006numerical}, which yields linear convergence and sensitivity to step size choices under ill-conditioning. Incorporating curvature via the performance Hessian can accelerate convergence to superlinear or quadratic rates. In reinforcement learning, this idea appears in natural policy gradients~\citep{amari1998,kakade2002natural}, trust-region methods~\citep{schulman2015trpo}, and quasi-Newton policy gradients~\citep{fazel2018global}.

A natural next step is to employ exact Newton methods that use the full Hessian of the performance function. However, computing the exact Hessian is typically challenging. It involves not only local curvature of the action-value function but also sensitivity of state distributions to the policy parameters. This distributional term couples the policy with the dynamics, and evaluating it requires differentiating through the transition kernel and the value function, which can be computationally expensive in general settings.

The linear quadratic regulator (LQR) provides a fundamental and analytically tractable setting for studying reinforcement learning algorithms~\citep{song2025convergencerobustnessvaluepolicy}. In this model, the dynamics are linear and the cost is quadratic, which ensures that the optimal policy is a linear state feedback. Crucially, in the LQR framework many quantities of interest, such as the value function, policy gradient, and even higher-order derivatives, admit closed-form expressions. This analytical structure makes it possible to compute not only the Gauss--Newton approximation but also the exact performance Hessian explicitly, thereby enabling efficient implementation of Newton-type methods that would be prohibitively costly in general reinforcement learning problems.

In this paper, we develop a second-order policy gradient framework for the LQR with known system matrices. We build on the recently developed performance Hessian theorem in~\cite{kordabad2022quasi}, formulated for general systems, and specialize it to the discounted LQR setting. This enables us to derive explicit closed-form expressions for the exact Hessian of the performance function. Moreover, we show that the approximate Hessian resulting from this formulation coincides with the Gauss--Newton structure studied in the literature~\citep{fazel2018global}. These closed-form characterizations provide curvature-aware updates that can be computed efficiently in LQR.

\textbf{Contributions.}
The main contributions of this paper are threefold: i) we prove that the Gauss--Newton (approximate) Hessian obtained from the general decomposition coincides with the classical LQR Gauss--Newton form; ii) we derive an explicit closed-form for the exact performance Hessian in discounted LQR under mild regularity assumptions, making exact Newton updates practical; and iii) we demonstrate on benchmarks that these second-order methods improve the convergence rate and stability over first-order policy gradient.

\textbf{Outline.}
The paper is organized as follows: Section~\ref{sec:preliminaries} introduces the discounted LQR setup, notation, and the recent second-order policy gradient theorem we build upon. Section~\ref{sec:grad-hess-lqr} derives closed-form expressions for the policy gradient, the Gauss--Newton Hessian, and the exact Hessian in the LQR setting. Section~\ref{sec:example} presents a scalar analytical example and numerical experiments illustrating the convergence properties of the proposed methods. Finally, Section~\ref{sec:conclusions} concludes and outlines directions for future work.

\textbf{Notation.}
We denote by $\mathbb{N}$, $\mathbb{N}_{\geq 0}$, and $\mathbb{R}$ the sets of positive integers, non-negative integers, and real numbers, respectively. For a symmetric matrix $A$, $A\succeq 0$ (resp.\ $A\succ0$) indicates positive semidefiniteness (resp.\ definiteness). We write $I_{m}$ for the $m\times m$ identity matrix. The normal vector to a set $\Omega\subseteq\mathbb{R}^n$ at $x\in\partial\Omega$ is denoted by $\boldsymbol n(x)$. The Frobenius and induced-2 norms are denoted by $\|\cdot\|_F$ and $\|\cdot\|$, respectively. For $ X \in \mathbb{R}^{m \times n} $, we define $ \mathrm{vec}(X)  \coloneqq  [X_1\tran , \dots, X_n\tran ]\tran  $, where $ X_i $ is the $ i $-th column of matrix $ X $. $\rho(X)$ denotes the spectral radius of $X$.
The Kronecker product is denoted by $\otimes$, where for $A\in\mathbb{R}^{m\times n}$ and $B\in\mathbb{R}^{p\times q}$ the matrix $A\otimes B\in\mathbb{R}^{mp\times nq}$ is defined entrywise by $(A\otimes B)_{(i-1)p+r,\,(j-1)q+s}=a_{ij}b_{rs}$ \citep{MatrixCookbook}.
The commutation matrix $K_{mn}\in\mathbb{R}^{mn\times mn}$ admits the explicit representation
\begin{equation*}
  K_{mn}:=\sum_{i=1}^{m}\sum_{j=1}^{n}\bigl(e_{m,i}\otimes e_{n,j}\bigr)\bigl(e_{n,j}\otimes e_{m,i}\bigr)\tran,
\end{equation*}
where $e_{m,i}\in\mathbb{R}^{m}$ and $e_{n,j}\in\mathbb{R}^{n}$ denote the $i$-th and $j$-th standard basis vectors, respectively (see e.g.,~\cite{Magnus1979}).

\section{Preliminaries and Background}
\label{sec:preliminaries}
In this paper, we consider discrete-time stochastic linear systems described by
\begin{equation}
  s_{k+1} = A s_k + B a_k + w_k, \quad k \in \mathbb{N}_{\geq 0},
  \label{eq:dynamics}
\end{equation}
where $s_k \in \mathbb{R}^n$ denotes the system state at time step $k$, $a_k \in \mathbb{R}^m$ is the control input and  $A\in\mathbb{R}^{n\times n}$, $B\in\mathbb{R}^{n\times m}$ are fixed known system matrices. The sequence $\{w_k\}^\infty_{k=0}$ represents independent and identically distributed (i.i.d.) random variables from a fixed distribution $w_k\sim p_w(\cdot)$ with zero mean and fixed covariance $\Sigma_w$, i.e., $\mathbb{E}[w_k]=0$ and $\mathbb{E}[w_k\,w_k \tran]=\Sigma_w$, whose support is contained in a known set $\mathcal{W}\subseteq\mathbb{R}^n$. The initial state $s_0$ is drawn from a known distribution $\rho_0$ with $\mathbb{E} [s_0 s_0\tran] = \Sigma_0$. 

In reinforcement learning, the dynamics are commonly represented by a transition kernel $p:\mathbb{R}^n\times \mathbb{R}^m \times \mathbb{R}^n\rightarrow [0,1]$. In particular, $p(s'| s,a)$ denotes the conditional distribution of the successor state $s'$ when a control input $a$ is applied to the system at state $s$. For the linear system in~\eqref{eq:dynamics}, this can be evaluated as follows:
\begin{align}
  p(s'| s,a) & = \int_{\mathcal{W}} \delta \big(s' - A s - B a - w\big)\, p_w(w)\,\mathrm{d}w \notag \\
             & = p_{w}\big(s' - A s - B a\big),
  \label{eq:Kernel}
\end{align}
where $\delta(\cdot)$ denotes the Dirac measure. The stage-wise cost $\ell:\mathbb{R}^n\times\mathbb{R}^m\rightarrow\mathbb{R}$ at a given state-input pair $(s,a)$ is given by the following quadratic function:
\begin{equation}
  \ell(s,a) = s\tran Q s + a\tran R a,
  \label{eq:stage-cost}
\end{equation}
where $Q \succeq 0$ and $R \succ 0$. A deterministic policy $\pi:\mathbb{R}^n\to\mathbb{R}^m$ maps each state $s$ to an input $a$. We consider a family of parametrized policies $\pi_\theta$. For the linear system~\eqref{eq:dynamics} with quadratic cost~\eqref{eq:stage-cost}, it is well-known that a linear state feedback policy of the form
\begin{align}
  \pi_\theta(s) \coloneqq -K s, \qquad K\in\mathbb{R}^{m\times n}, \label{eq:policy}
\end{align}
can be optimal for a cumulative stage-wise cost for a suitable choice of $K$. We parameterize the policy by $\theta \coloneqq  \mathrm{vec}(K)$.

The value function $V_\theta:\mathbb{R}^n\to\mathbb{R}$ corresponding to the linear policy \eqref{eq:policy} at a given state $s$ is  defined as the expectation of discounted infinite-horizon sum of the stage costs $\ell(s_k,a_k)$ under this policy, starting from the initial state $s_0=s$, i.e.,
\begin{equation*}
  V_\theta(s)
  =
  \mathbb{E}_{\tau_{\theta}}\left[
    \sum_{k=0}^{\infty}\gamma^{k}
    \ell(s_k, a_k)
    \middle| s_0 = s,\, a_k = -K s_k
    \right],
\end{equation*}
where $\gamma \in (0,1)$ is the discount factor. The expectation $\mathbb{E}_{\tau_{\theta}}[\cdot]$ is taken over the state-action trajectory generated by the underlying system dynamics~\eqref{eq:dynamics}, and the policy.
The action-value function $Q_\theta: \mathbb{R}^n\times \mathbb{R}^m\rightarrow \mathbb{R}$ is also defined using the Bellman equation as
\begin{equation*}
  Q_\theta(s,a)  =\ell(s,a)+\gamma\,\mathbb{E}_{w}\bigl[V_{\theta}(s')\,|\, s, a\bigr].
\end{equation*}

In the following, we introduce stabilization notions for the described discounted LQR problem that enable us to provide well-defined value functions  for a given linear control policy.
\begin{Definition}[Stabilization notions]
  \label{Def:Stabilization}
  For the linear system \eqref{eq:dynamics}, the linear policy \eqref{eq:policy} is \emph{$\gamma$-stabilizing} if $\rho(\sqrt{\gamma}A_{\theta})<1$, where $A_{\theta} \coloneqq A-BK$ is the closed-loop matrix. Moreover, the pair $(A,B)$ is \emph{$\gamma$-stabilizable} if there exists a $\gamma$-stabilizing policy in the form of $\pi_\theta(s) = -K s$.
\end{Definition}

Note that if a policy is $\gamma$-stabilizing, then it is also $\gamma_0$-stabilizing for all $\gamma_0 \leq \gamma$, while the reverse may not hold. Moreover, any policy can be $\gamma$-stabilizing for sufficiently small $\gamma$. Therefore, requiring a policy to be $\gamma$-stabilizing is a more relaxed condition than requiring it to be $1$-stabilizing, which corresponds to the classical notion of stability without a discount factor. Introducing a proper discount factor thus relaxes the classical stability requirement.

Now we provide a lemma that gives explicit expressions for the value functions of LQR in the discounted setting.

\begin{Lemma} \label{lem:stabilization}
  For the linear system~\eqref{eq:dynamics} with quadratic stage cost~\eqref{eq:stage-cost}, the value function $V_\theta$ and action-value function $Q_\theta$ corresponding to a $\gamma$-stabilizing linear policy $\pi_\theta(s) = -K s$ are obtained as follows:
  \begin{subequations}
    \begin{align}
      V_{\theta}(s) & = s\tran P_{\theta}s+q_{\theta},\label{eq:V-theta}                                            \\
      Q_\theta(s,a) & = s\tran \bigl(Q+\gamma A\tran P_{\theta}A\bigr)s+2 \gamma s\tran  A\tran P_{\theta}B a\notag \\& \quad \quad + a\tran \bigl(R+\gamma B\tran P_{\theta}B\bigr)a +q_{\theta}.
      \label{eq:Q-theta}
    \end{align}
  \end{subequations}
  where
  \begin{subequations}
    \begin{align}
      P_{\theta} & =Q+K\tran RK   +\gamma\,A_{\theta}\tran P_{\theta}A_{\theta},\label{eq:discrete-Lyap} \\
      q_{\theta} & =\frac{\gamma}{1-\gamma}\;
      \tr(P_{\theta} \Sigma_w).
      \label{eq:q-solution}
    \end{align}
  \end{subequations}
\end{Lemma}
\noindent The result for \eqref{eq:V-theta} and \eqref{eq:Q-theta} follows from the discounted Bellman equation by adapting the classical undiscounted LQR derivations \citep{anderson2007optimal,bertsekas1995dynamic} to the discounted setting. In infinite-horizon stochastic LQR with additive noise, a finite value requires $\gamma<1$, since \eqref{eq:q-solution} diverges at $\gamma=1$.

The performance (or objective) function $J(\theta)$ for policy parameter $\theta$ is defined as the expected value of the value function $V_\theta$ under the initial state distribution $\rho_0$, i.e.,
\begin{align}
  J(\theta)   \coloneqq  \mathbb{E}_{s_0\sim\rho_0}[V_{\theta}(s_0)].
\end{align}
The optimal value function and the optimal parameter vector are defined as follows:
\begin{equation*}
  V^\star(s)=\min_{\theta} V_\theta( s),\qquad \theta^\star\in \mathrm{arg}\min_{\theta} J(\theta). \notag
\end{equation*}

Under the standard assumption that $(A,B)$ is $\gamma$-stabilizable in Definition~\ref{Def:Stabilization}, there exists an optimal policy of the form \eqref{eq:policy}. The optimal linear feedback gain is
\begin{equation}
  K^\star = \bigl(R + \gamma B \tran P^\star B\bigr)^{-1}\, \gamma B \tran P^\star A,
  \label{eq:gain-theta}
\end{equation}
see, e.g., \cite{anderson2007optimal,bertsekas1995dynamic,tedrake2025lqr}. The corresponding optimal value function is $V^\star(s)=s \tran P^\star s + q^\star$, where $P^\star$ and $q^\star$ are obtained by substituting $K=K^\star$ into \eqref{eq:discrete-Lyap} and \eqref{eq:q-solution}.

At first sight, \eqref{eq:gain-theta} appears to resolve the optimal control problem, but it is an implicit characterization: it depends on $P^\star$, the solution of the discounted Lyapunov equation \eqref{eq:discrete-Lyap}. Thus, computing $K^\star$ still requires solving that matrix equation. Even with known $(A,B,Q,R)$, direct policy optimization is meaningful; as $J(\theta)$ is nonconvex and finite only on the stabilizing set, yet it satisfies a global convergence of first-order methods under suitable step sizes \citep{karimi2020linearconvergencegradientproximalgradient,bhandari2022globaloptimalityguaranteespolicy}.

These considerations motivate preconditioned, gradient-based policy optimization. A generic update is
\begin{align}
  \theta_{k+1} = \theta_k - \alpha_k\, P_k^{-1}\, \nabla_\theta J(\theta)\big|_{\theta=\theta_k}, \label{eq:preconditioned-step}
\end{align}
where $P_k\succ 0$ is a preconditioner and $\alpha_k>0$ is a step size. Standard choices are: $P_k=I$, the common first-order policy gradient method; $P_k= F_k$, where $F_k$ is the Fisher information matrix, the natural gradient; and $P_k=\nabla_\theta^2 J(\theta_k)$, Newton's method. Because exact Hessians are expensive and often nontrivial in RL, one often uses $P_k=H_k$ with a tractable surrogate $H_k\approx\nabla_\theta^2 J(\theta_k)$ (e.g., Gauss--Newton). Under standard smoothness and stabilizability assumptions, such quasi-Newton choices retain strong local convergence while avoiding full second-order computation cost~\citep{furmston2016approximate,fazel2018global}.

We next recall the deterministic policy gradient theorem and its Hessian extension. The following assumptions ensure that the performance function $J(\theta)$ is twice differentiable.

\begin{Assumption}[Regularity]\label{ass:regularity}
  The maps $p(s'|s,a)$, $\pi_\theta(s)$, $\ell(s,a)$, $\rho_0(s)$ and the derivatives $\nabla_a p(s'|s,a)$, $\nabla_\theta \pi_\theta(s)$, $\nabla_a \ell(s,a)$, $\nabla_a^2 p(s'|s,a)$, $\nabla_\theta^2 \pi_\theta(s)$, $\nabla_a^2 \ell(s,a)$ are continuous in their arguments. $\rho_0$ and $p$ are uniformly bounded, and $\nabla_a \ell$, $\nabla_a p$, $\nabla_a^2 p$, $\nabla_a^2 \ell$ are uniformly bounded for some constants.
\end{Assumption}
\noindent One can verify that for discounted LQR with dynamics~\eqref{eq:dynamics}, cost~\eqref{eq:stage-cost}, and linear policy~\eqref{eq:policy}, the regularity conditions assumptions in Assumption~\ref{ass:regularity} hold automatically. We next state the policy-gradient theorem and the associated Hessian decomposition.

\begin{theorem}[Policy Gradient and Hessian]
  \label{thm:DPG-both}
  For dynamics with transition kernel~\eqref{eq:Kernel} and under Assumption~\ref{ass:regularity}, the policy gradient and Hessian admit the following expressions:
  \begin{subequations}
    \begin{align}
      \nabla_{\theta}J(\theta)
       & =
      \mathbb{E}_{\tau_{\theta}} \Big[
      \nabla_{\theta}\pi_{\theta}(s)\,
      \nabla_{a}Q_{\theta}(s,a)\,\big|_{\,a=\pi_{\theta}(s)}
      \Big],
      \label{eq:DPG-gradient} \\
      \nabla_{\theta}^{2}J(\theta)
       & =
      H(\theta)+\gamma\,\Lambda(\theta),
      \label{eq:DPG-hessian}
    \end{align}
  \end{subequations}
  where
  \begin{subequations}
    \begin{align}
      H(\theta)       \coloneqq \mathbb{E}_{\tau_\theta}\Big[ & \nabla_{\theta}^{2}\pi_{\theta}(s)\otimes \nabla_{a}Q_{\theta}(s,a)\Big|_{a=\pi_{\theta}(s)}+ \nonumber       \\
                      & \quad \nabla_{\theta}\pi_{\theta}(s)\nabla_{a}^{2}Q_{\theta}(s,a)\Big|_{a=\pi_{\theta}(s)}\nabla_{\theta}\pi_{\theta}(s)\tran \Big],
      \label{eq:H-policy-hessian}                                                                                                                                         \\[4pt]
      \Lambda(\theta) \coloneqq \mathbb{E}_{\tau_\theta}\Big[ & \int\nabla_{\theta}p(s'|s,\pi_{\theta}(s))\nabla_{\theta}V_{\theta}(s')\tran \mathrm{d}s'+ \nonumber \\
                      & \quad \int\nabla_{\theta}V_{\theta}(s')\nabla_{\theta}p(s'|s,\pi_{\theta}(s))\tran \mathrm{d}s'\Big].
      \label{eq:Lambda-policy-hessian}
    \end{align}
  \end{subequations}
\end{theorem}

\begin{pf}
  Under Assumption~\ref{ass:regularity}, the deterministic policy gradient theorem \citep{silver2014deterministic}, \eqref{eq:DPG-gradient} holds. Differentiating \eqref{eq:DPG-gradient} yields \eqref{eq:DPG-hessian} together with \eqref{eq:H-policy-hessian}–\eqref{eq:Lambda-policy-hessian}; see \cite{kordabad2022quasi} for the derivation. \hfill $\blacksquare$
\end{pf}

Note that Theorem~\ref{thm:DPG-both} applies to general dynamics. The performance Hessian decomposes as \eqref{eq:DPG-hessian}, where $H(\theta)$ is a tractable curvature term and $\Lambda(\theta)$ captures distributional effects. The surrogate $H(\theta)$ is readily computable and is exact at the optimum, i.e., $H(\theta^\star)=\nabla_\theta^2 J(\theta^\star)$ (see Theorem 3 in \cite{kordabad2022quasi}), which justifies Gauss--Newton updates. In contrast, $\Lambda(\theta)$ is typically costly to evaluate because it depends on the transition kernel.

\section{Gradient and Hessian for LQR}
\label{sec:grad-hess-lqr}
A central contribution of this work is the evaluation of the general policy Hessian framework stated in Theorem~\ref{thm:DPG-both} for the LQR problem. Leveraging the analytical structure of the LQR, we obtain an explicit, structured, and computationally tractable representation of the second-order policy gradient. This explicit characterization goes beyond the abstract formulations available in the general setting and provides novel theoretical insights into the geometry of the policy optimization landscape. We first state the differential identities for $Q_\theta$ and $\pi_\theta$ required in Theorem~\ref{thm:DPG-both}. For $Q_\theta$,
\begin{subequations}
  \begin{align}
                & \nabla_{a} Q_\theta(s,a)
    = 2\big(R + \gamma\,B\tran P_\theta B\big)a
    + 2\gamma\,B\tran P_\theta A\,s, \notag                                                                       \\
    \Rightarrow & \left.\nabla_{a} Q_\theta(s,a)\right|_{a=-Ks}
    = -2\big(RK - \gamma\,B\tran P_\theta A_\theta\big)s,\label{eq:Q-grads-main}                                  \\
    \Rightarrow & \,\,\nabla_{a}^2 Q_\theta(s,a)= 2\big(R + \gamma\,B\tran P_\theta B\big),\label{eq:Q-hess-main}
  \end{align}
\end{subequations}

and for the linear policy \eqref{eq:policy}, we obtain
\begin{equation}\label{eq:policy-grad-main}
  \nabla_{\theta}\pi_\theta(s) = -\,s\otimes I_m, \qquad \nabla_{\theta}^{2}\pi_\theta(s)
  = 0.
\end{equation}
These equations provide all ingredients for substitution into~\eqref{eq:H-policy-hessian}--\eqref{eq:Lambda-policy-hessian}.

\subsection{Policy Gradient in LQR}
We now evaluate the policy gradient~\eqref{eq:DPG-gradient} for the LQR setting. Substituting the action and policy derivatives from \eqref{eq:Q-grads-main} and \eqref{eq:policy-grad-main} yields
\begin{align}
  \nabla_{\theta}J(\theta)
   & = \mathbb{E}_{\tau_{\theta}} \Big[\nabla_{\theta} \pi_{\theta}(s)\, \nabla_{a}Q_\theta(s,a)\big|_{a=-K s}\Big] \notag           \\[4pt]
   & = 2\,\mathbb{E}_{\tau_{\theta}} \Big[(s\otimes I_m)\big(RK-\gamma B\tran P_{\theta} A_{\theta}\big)s\Big] \notag                \\[4pt]
   & = 2\,\mathbb{E}_{\tau_{\theta}} \Big[\mathrm{vec} \Big(\big(RK-\gamma B\tran P_{\theta} A_{\theta}\big)ss\tran\Big)\Big] \notag \\[4pt]
   & = 2\,\mathrm{vec} \Big(\big(RK-\gamma B\tran P_{\theta} A_{\theta}\big)\,\Sigma_\theta\Big),
  \label{eq:J-grad-vec}
\end{align}
where we use $\mathrm{vec}(v s\tran)=(s\otimes I_m)v$ in the third equality.

Matrix $\Sigma_\theta$ is the discounted state second-moment matrix under policy $\pi_\theta$ and is defined as,

\begin{equation}
  \Sigma_\theta \coloneqq \mathbb{E}_{\tau_\theta}[s s\tran]
  = \sum_{k=0}^\infty \gamma^{k}\,\mathbb{E}[s_k s_k\tran].
  \label{eq:sigma-theta}
\end{equation}
This is the discounted state correlation matrix \citep{pmlr-v139-wang21j}, also called the discounted state-occupancy measure \citep{bhandari2022globaloptimalityguaranteespolicy}. The next lemma gives a closed form characterization of $\Sigma_\theta$ for discounted LQR.

\begin{Lemma}[Discounted State Correlation Matrix]
  \label{thm:second-moments}
  For any $\gamma$-stabilizing gain $K$ in Definition~\ref{Def:Stabilization}, the series in \eqref{eq:sigma-theta} converges and $\Sigma_\theta$ satisfies the discounted Lyapunov equation
  \begin{equation}
    \Sigma_\theta - \gamma\,A_\theta\,\Sigma_\theta\,A_\theta\tran
    = \Sigma_0 + \frac{\gamma}{1-\gamma}\,\Sigma_w .
    \label{eq:discounted-lyap-sigma}
  \end{equation}
\end{Lemma}
\begin{pf}
  See Appendix~\ref{app:second-moments}.  \hfill $\blacksquare$
\end{pf}

The covariance matrix $\Sigma_\theta$ appears explicitly in the policy gradient expression~\eqref{eq:J-grad-vec} and, as will be shown subsequently, also arises in both the exact and quasi-Hessian formulations.

\subsection{Quasi-Newton Policy Gradient in LQR}
We derive an approximation of the Hessian of the performance function for discounted LQR by adapting the \emph{general} second-order formulations \eqref{eq:H-policy-hessian}. From \eqref{eq:H-policy-hessian}, and since $\nabla_{\theta}^{2}\pi_{\theta}(s)=0$ for linear policies, the first term vanishes. Using \eqref{eq:Q-hess-main} and \eqref{eq:policy-grad-main}, we can rewrite $H(\theta)$ as follows:

\begin{align}
  H(\theta)
   & =
  \mathbb{E}_{\tau_{\theta}}\Bigl[
    \nabla_{\theta}\pi_{\theta}(s)\,
    \nabla_{a}^{2}Q_\theta(s,a)\bigl\rvert_{a=\pi_{\theta}(s)}
    \nabla_{\theta}\pi_{\theta}(s)\tran
  \Bigr] \notag                                                                                             \\
   & = 2 \,
  \mathbb{E}_{\tau_{\theta}}\Bigl[
    (s\otimes I_m)
    \bigl(R+\gamma B\tran P_{\theta}B\bigr)
    (s\tran \otimes I_m)
  \Bigr]\notag                                                                                              \\
  %  & = 2 \,
  % \mathbb{E}_{\tau_{\theta}}\Bigl[
  %   (s\otimes I_m)
  %   \Big(s\tran \otimes (R+\gamma B\tran P_{\theta}B) \Big)
  % \Bigr] \notag                                                                                             \\
   & = 2 \, \mathbb{E}_{\tau_{\theta}}\Bigl[(s\,s\tran )\,\otimes\,
    \bigl(R+\gamma B\tran P_{\theta}B\bigr)
  \Bigr] \notag                                                                                             \\
   & = 2 \, \Sigma_\theta \otimes \bigl(R+\gamma B\tran P_{\theta}B\bigr), \label{eq:GN-Hessian-estimation}
\end{align}
using the Kronecker mixed-product rule.

\begin{Remark}
  Most Gauss--Newton LQR derivations treat deterministic, \emph{undiscounted} LQR~\citep{Li2004IterativeLQ,Giftthaler18}. Here we instantiate~\eqref{eq:H-policy-hessian} for $\gamma\in(0,1)$ with nonzero additive noise. When $\gamma\to1$, applying~\eqref{eq:GN-Hessian-estimation} to~\eqref{eq:J-grad-vec} recovers the classical update \cite[Equation~(7)]{fazel2018global} along with standard convergence guarantees. Thus, the treatment both extends discounted stochastic LQR and unifies with established formulations.
\end{Remark}

\subsection{Exact Hessian in LQR}
To leverage second-order methods beyond Gauss--Newton, one must capture the state distribution gradients with respect to the policy parameters in~\eqref{eq:DPG-hessian}. In general RL this coupling is what makes exact Newton steps impractical. Our key insight is that, for discounted LQR, this coupling admits an explicit form with modest assumptions on the disturbance.

We begin with a mild regularity requirement ensuring that boundary contributions vanish when differentiating through the transition kernel. This condition is satisfied by essentially most of the distributions used in control, including Gaussian and sub-Gaussian distributions.
\begin{Assumption}[Vanishing boundary flux]\label{assump:vanishing-boundary-flux}
  Assume one of the following holds:
  \begin{enumerate}[label=(\roman*)]
    \item Bounded support: $\mathcal{W}$ is a Lipschitz domain and $p_w(w)=0$ for all $w\in\partial\mathcal{W}$; or
    \item Unbounded support: $\mathcal{W}=\mathbb{R}^n$ and
          \begin{align}
            \lim_{R\to\infty} R^{n+1}\,\sup_{\|w\|=R} p_w(w)=0. \label{eq:unbounded-support}
          \end{align}
  \end{enumerate}
\end{Assumption}
\noindent Under Assumption~\ref{assump:vanishing-boundary-flux}, boundary terms vanish, yielding a closed-form transition contribution to the policy Hessian. The assumption holds for standard disturbances (see remark~\ref{rem:discussion-bounded-i}). With this assumption and the regularity conditions of Theorem~\ref{thm:DPG-both}, the exact Hessian admits a closed form in LQR.
\begin{theorem}[Exact Hessian in LQR]\label{thm:Lambda_structure}
  Under Assumption~\ref{assump:vanishing-boundary-flux}, the exact Hessian of the performance function for LQR is obtained as
  $\nabla_{\theta}^{2}J(\theta)=H(\theta)+\gamma\,\Lambda(\theta)$
  where $H(\theta)$ is evaluated from~\eqref{eq:GN-Hessian-estimation} and
  \begin{align}
    \Lambda(\theta) & = -2 \bigg[(\Sigma_\theta A_\theta \tran\otimes B \tran) \frac{\partial\,\mathrm{vec}(P_\theta)}{\partial\theta} \notag \\ &\qquad \qquad \qquad\qquad +\frac{\partial\,\mathrm{vec}(P_\theta)}{\partial\theta}\tran (A_\theta \Sigma_\theta \otimes B)\bigg], \label{eq:Lambda-final}
  \end{align}
  and where the Jacobian of $P_\theta$ with respect to $\theta$ is
  \begin{align}
    \frac{\partial\,\mathrm{vec}(P_\theta)}{\partial \theta} = T_\theta^{-1} \Big[(S_\theta\tran \otimes I_n) \,K_{mn} + (I_n \otimes  S_\theta\tran)\Big], \label{eq:J-grad-vec-P}
  \end{align}
  with
  \begin{align}
    S_\theta \coloneqq R K - \gamma B\tran P_\theta A_\theta,\quad
    T_\theta \coloneqq I_{n^2} - \gamma\,(A_\theta\tran \otimes A_\theta\tran). \label{eq:S-and-Iinv}
  \end{align}
\end{theorem}
\begin{pf}
  See Appendix~\ref{app:proof-Lambda-structure}.  \hfill $\blacksquare$
\end{pf}
The term $\Lambda(\theta)$ captures the second-order sensitivity of the value-function gradient to policy-induced changes in the transition kernel via its coupling with $\nabla_\theta V_{\theta}(s')$. The representation in Theorem~\ref{thm:Lambda_structure} enables efficient evaluation in exact Newton methods for LQR, which has second-order global convergence. To compute the exact Hessian, evaluate the gradient in \eqref{eq:J-grad-vec}; solve \eqref{eq:discrete-Lyap} for $P_\theta$ and \eqref{eq:discounted-lyap-sigma} for $\Sigma_\theta$; build the Jacobian \eqref{eq:J-grad-vec-P} using \eqref{eq:S-and-Iinv}; assemble $H(\theta)$ via \eqref{eq:GN-Hessian-estimation} and $\Lambda (\theta)$ via \eqref{eq:Lambda-final}.

Therefore, Theorem~\ref{thm:Lambda_structure} extends Gauss–Newton to exact Newton treatments by providing a closed-form expression for the exact Hessian, thereby enabling true Newton steps for stochastic LQR that remain computationally practical.

\section{Analytical Example and Simulations}\label{sec:example}
We first analyze a one-dimensional instance to validate the derivations. We next present numerical experiments on two benchmarks: an inverted-pendulum linearization and a high-dimensional seismic shear-building model, both with strongly anisotropic objective landscapes. Numerical results demonstrate the advantages of exact second-order information for Newton-type policy optimization, with and without line search.
\subsection{Analytical example: scalar LQR}
Consider the following scalar discounted LQR
\begin{align}
  s_{k+1}=as_k+ba_k+w_k,\notag
\end{align}
with $a_k = - \theta s_k$, $w_k\sim\mathcal{N}(0,\sigma^2)$ and $s_0\sim\mathcal{N}(0,\sigma_0^2)$. The resulting closed-loop dynamics are $s_{k+1}=A_{\theta} s_k + w_k$. If $\theta$ is $\gamma$-stabilizing, then using~\eqref{eq:discrete-Lyap} and~\eqref{eq:sigma-theta}, we obtain the closed-form expressions
\begin{align}
  \Sigma_{\theta} = \frac{\sigma_0^{2} + \dfrac{\gamma}{1-\gamma}\,\sigma^{2}}{1-\gamma\,(A_{\theta})^{2}}, \qquad
  P_\theta = \frac{Q + R\,\theta^{2}}{1-\gamma\,(A_{\theta})^{2}}. \notag
\end{align}
Invoking \eqref{eq:J-grad-vec-P}, the derivative of $P_\theta$ with respect to $\theta$ is
\begin{align}
  \frac{\partial P_\theta}{\partial \theta} = \frac{2\,S_\theta}{1 - \gamma\,A_\theta^{2}}, \quad   S_\theta \coloneqq R\,\theta - \gamma\,b\,P_\theta\,A_\theta. \notag
\end{align}
Using \eqref{eq:J-grad-vec}, the policy gradient in one dimension is
\begin{align}
  \frac{\partial J(\theta)}{\partial \theta}
   & = 2\,\Sigma_\theta\,(R\,\theta - \gamma\,b\,P_\theta\,A_{\theta}). \label{eq:1d-J-grad}
\end{align}
Likewise, by \eqref{eq:GN-Hessian-estimation}, the quasi-Hessian in one dimension is
\begin{align}
  H(\theta) & = 2\,\Sigma_\theta\,(R + \gamma\,b^2 P_\theta).\label{eq:1d-GN-Hessian-estimation}
\end{align}
Using \eqref{eq:Lambda-final}, the transition contribution to the Hessian is
\begin{align}
  \Lambda(\theta) = -4\,\Sigma_\theta\,A_\theta\,b \,\frac{\partial P_\theta}{\partial \theta}. \label{eq:Lambda-1d}
\end{align}

Moreover, differentiate \eqref{eq:1d-J-grad} once more to obtain the exact Hessian analytically
\begin{align}
  \frac{\partial^2 J(\theta)}{\partial \theta^2} = 2 \Big( S'_\theta \Sigma_\theta + S_\theta \Sigma'_\theta \Big), \label{eq:1d-J-hess}
\end{align}
where
\begin{gather}
  \Sigma'_\theta =-\frac{2\gamma b\,A_\theta}{\,1-\gamma A_\theta^2\,}\,\Sigma_\theta =-\frac{\gamma b\,A_\theta}{S_\theta}\,\Sigma_\theta\, \frac{\partial P_\theta}{\partial\theta}, \notag \\
  S'_\theta=R-\gamma b\,A_\theta\frac{\partial P_\theta}{\partial\theta}+\gamma b^2 P_\theta \notag.
\end{gather}
Hence \eqref{eq:1d-J-hess} becomes
\begin{align}
  \frac{\partial^2 J(\theta)}{\partial\theta^2} & =2\,\Sigma_\theta\bigl(R+\gamma b^2 P_\theta\bigr)-4\gamma b\,\Sigma_\theta\frac{\partial P_\theta}{\partial\theta}\,A_\theta \notag \\
                                                & =H(\theta)+\gamma \Lambda(\theta), \notag
\end{align}
where $H(\theta)$ and $\Lambda(\theta)$ are given in \eqref{eq:1d-GN-Hessian-estimation} and \eqref{eq:Lambda-1d}, respectively. This confirms that the derived expressions for $H(\theta)$ and $\Lambda(\theta)$ correctly recover the exact Hessian in one dimension.
One can verify that the result in \cite{kordabad2022quasi} is a special case of this derivation with $a=b=1$ and $Q=R=0.5$.

\subsection{Inverted Pendulum Control}
\begin{figure}[t!]
  \centering
  \includegraphics[width=1\linewidth]{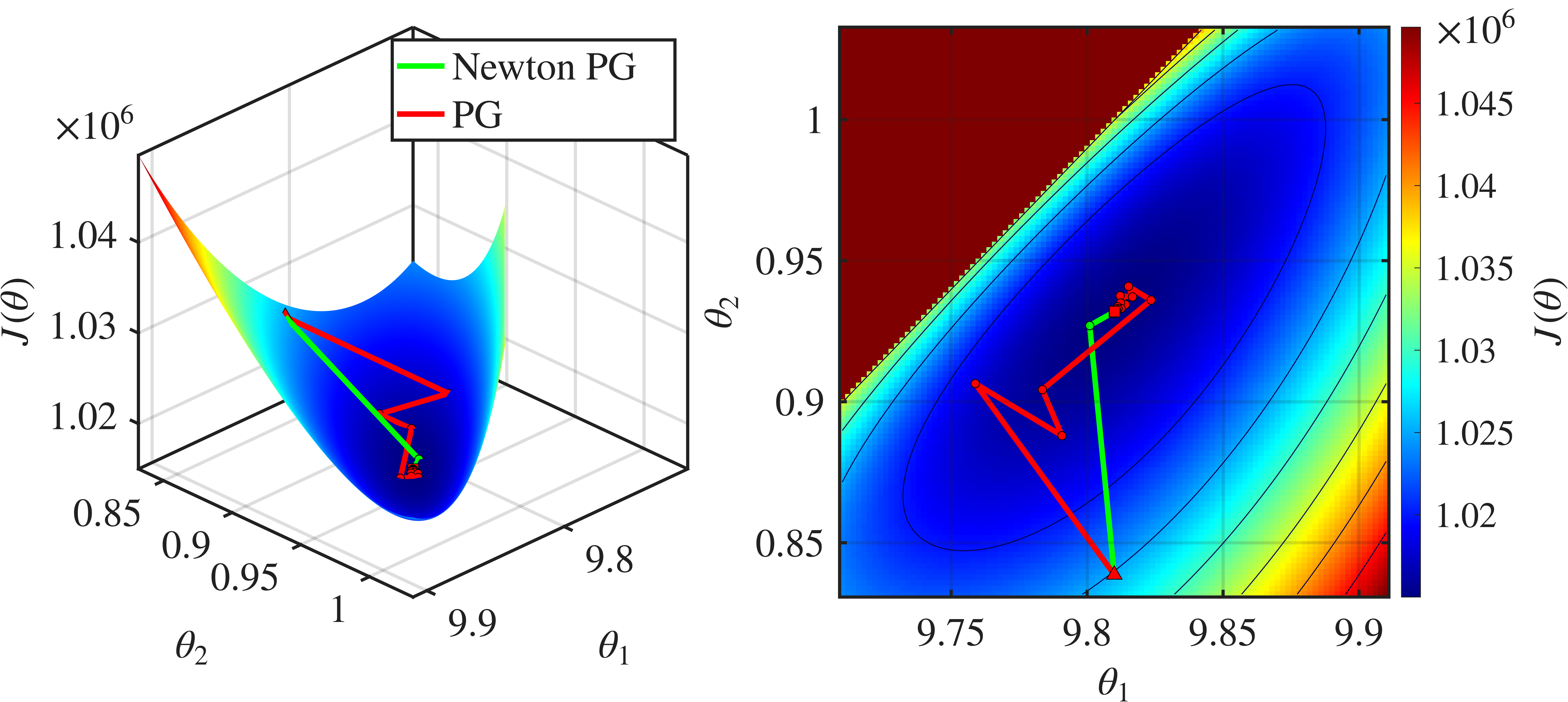}
  \caption{\textbf{Left:} 3-D surface of the cost function $J(\theta)$ with trajectories. \textbf{Right:} Heatmap of the same region. The Newton PG (green) proceeds directly toward the optimal parameters, whereas the first order PG (red) fluctuates. Darker blue indicates lower cost.}
  \label{fig:jk_landscape}
\end{figure}

We study the discretization of the upright linearization of a planar inverted pendulum. The dynamics are in the form of~\eqref{eq:dynamics} with $A$ and $B$ given as
\begin{align}
  A & = \begin{bmatrix}
          0   & 1 \\
          g/l & 0
        \end{bmatrix}, \quad B = \begin{bmatrix}
                                   0 \\
                                   1/(m l^2)
                                 \end{bmatrix}, \notag
\end{align}
where $g=9.81\,\mathrm{m/s^2}$ is the gravitational constant, $l=1\,\mathrm{m}$ is the pendulum length, and $m=1\,\mathrm{kg}$ is the mass.
Process noise is i.i.d. Gaussian with covariance $\Sigma_w = I_2$, and the initial state covariance is $\Sigma_0 = 0.1\,I_2$. Performance is evaluated under an infinite-horizon discounted cost with discount factor $\gamma=0.9$.

The state penalty is strongly anisotropic with eigenvalues $(\lambda_1,\lambda_2)=(10^5,10^{-4})$ rotated by $\psi=40^\circ$, implemented via $Q = C \, \mathrm{diag}(\lambda_1,\lambda_2)\, C\tran $ where $C$ is the rotation matrix. The input penalty is $R=0.1$. Policy gradient schemes are initialized at a common stabilizing gain $K_0$ computed with \texttt{dlqr}. Step sizes are selected by backtracking line search~\citep{nocedal2006numerical}. Figure~\ref{fig:jk_landscape} displays the discounted LQR objective $J(\theta)$ with parameters $(\theta_1,\theta_2)$ slice with the corresponding trajectories. Newton follows the rotated, anisotropic valley to the optimal solution in a few steps, whereas the first order policy gradient oscillates, highlighting the benefit of curvature information. Curvature-based preconditioning aligns updates with principal directions, yielding larger per-iteration decreases in $J(\theta)$ and more stable iterates.

\subsection{Seismic Shear-Building Benchmark}
\begin{figure}[t]
  \centering
  \includegraphics[width=1\linewidth]{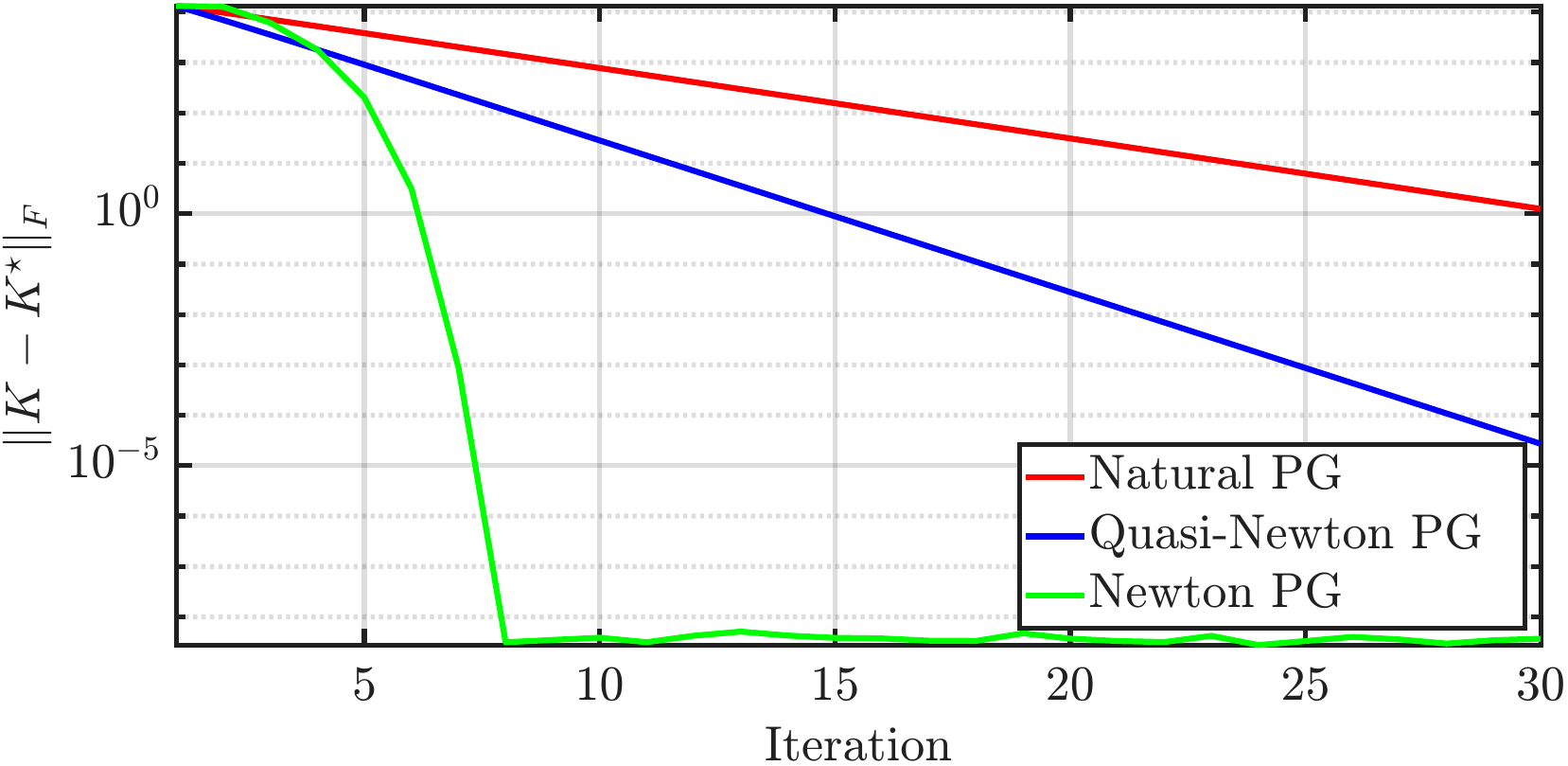}
  \caption{The Frobenius-norm policy error, $\|K_k - K^\star\|_F$, versus iteration index $k$ for natural policy gradient, Gauss--Newton, and Newton. All algorithms are initialized at the same stabilizing gain $K_0$.}
  \label{fig:building_convergence}
\end{figure}

We evaluate a multi-story shear-building benchmark under base excitation. The stacked state is $s_k=[\,q_k\tran,\ \dot q_k\tran\,]\tran\in\mathbb{R}^{48}$ with interstory displacements $q_k\in\mathbb{R}^{24}$ and velocities $\dot q_k\in\mathbb{R}^{24}$. The control input $a_k$ applies base actuation. A discrete-time model of the form~\eqref{eq:dynamics} is obtained by first-order augmentation and zero-order-hold discretization with $T_s=0.01\,\mathrm{s}$; see~\cite{morAntSG01} for the resulting matrices. We set $w_k \sim \mathcal{N}(0,10^{-4} I_{48})$ and $s_0 \sim \mathcal{N}(0,10^{-2} I_{48})$ and neglect measurement noise. The state penalty is strongly anisotropic,
\begin{align}
  Q = V\,\mathrm{diag} \big(\lambda_{\mathrm{hi}} I_k, \lambda_{\mathrm{lo}} I_{\,n-k}\big)V\tran + \varepsilon I,\quad ,\varepsilon>0, \notag
\end{align}
where $V\in\mathbb{R}^{48\times 48}$ is a  orthogonal basis, $0<\lambda_{\mathrm{lo}}\ll\lambda_{\mathrm{hi}}$ and $R=0.01$. The initial stabilizing gain $K_0$ is computed using \texttt{dlqr}. Step sizes are set to $\alpha_{GN}=0.5$ for Gauss--Newton, in accordance with convergence results in \cite{fazel2018global}, and to $\alpha_N=1$ for Newton, selected by a single tuning pass and then kept constant. Figure~\ref{fig:building_convergence} reports $\|K_k-K^\star\|_F$ as a function of the iteration index. The plot indicates that Newton achieves quadratic local convergence, while Gauss--Newton attains superlinear rates, both substantially outperforming first-order policy gradient. 
\section{Conclusions}\label{sec:conclusions}
We presented a curvature-aware policy optimization framework for discounted stochastic LQR that yields explicit formulas for both the Gauss--Newton surrogate and the exact performance Hessian. The surrogate coincides with the classical LQR Gauss--Newton matrix; the exact Hessian augments it with a distributional term evaluable under mild boundary conditions. Future work includes model-free curvature estimation (actor--critic, off-policy) with finite-sample guarantees and robust under model uncertainty.
\bibliography{QRL}

\appendix
\section{Proof of Lemma~\ref{thm:second-moments}}
\label{app:second-moments}
\begin{pf}
  First, the one-step second-moment recursion follows from the dynamics and the independence and zero-mean of $w_k$:
  \begin{equation}\label{eq:one-step}
    \mathbb{E}[s_{k+1}s_{k+1}\tran] = A_\theta\,\mathbb{E}[s_k s_k\tran]\,A_\theta\tran+\Sigma_w, \qquad k\ge 0,
  \end{equation}
  because the cross terms $\mathbb{E}[A_\theta s_k w_k\tran]$ and $\mathbb{E}[w_k s_k\tran A_\theta\tran]$ vanish. By definition in \eqref{eq:sigma-theta}, we have
  \begin{align*}
    \Sigma_\theta-&\gamma A_\theta\Sigma_\theta A_\theta\tran  =\sum_{k=0}^\infty \gamma^k \mathbb{E}[s_k s_k\tran] -\sum_{k=0}^\infty \gamma^{k+1} A_\theta \mathbb{E}[s_k s_k\tran] A_\theta\tran \notag \\
                                                             & =\Sigma_0 +\sum_{k=1}^\infty \gamma^k\big(\mathbb{E}[s_k s_k\tran]-A_\theta \mathbb{E}[s_{k-1}s_{k-1}\tran]A_\theta\tran\big) \notag        \\
                                                             & \overset{\eqref{eq:one-step}}{=}\Sigma_0 + \sum_{k=1}^\infty \gamma^k\,\Sigma_w  = \Sigma_0 + \frac{\gamma}{1-\gamma}\,\Sigma_w. \hspace{1.8cm} \blacksquare
  \end{align*}
\end{pf}

\section{Proof of Theorem~\ref{thm:Lambda_structure}}
\label{app:proof-Lambda-structure}
Theorem~\ref{thm:Lambda_structure} follows from two lemmas: (i) boundary terms vanish when integrating the disturbance density by parts; (ii) the Jacobian of the discounted Lyapunov solution with respect to the policy is characterized. We then combine them to complete the proof.
\begin{Lemma}[Vanishing boundary flux]\label{thm:vanishing-boundary-flux}
  Under Assumption~\ref{assump:vanishing-boundary-flux}, the boundary flux of any at-most-quadratic vector field $g$ through the support boundary of $p_w$ vanishes:
  \begin{align}
    \int_{\partial\mathcal W} p_w(w)\, \boldsymbol n(w) g(A_{\theta} s+w) \tran \,\mathrm{d} S = 0, \notag
  \end{align}
  where $\boldsymbol n(w)$ is the outward unit normal on the boundary $\partial\mathcal W$.  
\end{Lemma}
\begin{pf}
  We treat the two cases in Assumption~\ref{assump:vanishing-boundary-flux}.

  \emph{Case (i): $\mathcal W$ bounded and $p_w=0$ on $\partial\mathcal W$.}
  Since $p_w$ vanishes on the boundary and $\|\boldsymbol n(w) g(A_{\theta} s+w) \tran\| < \infty$ for some $w < \infty$, the integrand is identically zero and hence the integral is zero.

  \emph{Case (ii): $\mathcal W=\mathbb R^n$ with tail decay.}
  For $R>0$ let $S_R \coloneqq \{w\in\mathbb R^n:\|w\|=R\}$ and define
  \begin{align}
    B_R  \coloneqq  \int_{S_R} p_w(w)\, \boldsymbol n(w) \, g(A_\theta s+w) \tran \, \mathrm{d} S  . \notag
  \end{align}
  Because $g$ is quadratic, there exists $C_0>0$ such that
  \begin{align}
    \|g(x)\|\le C_0\bigl(1+\|x\|^2\bigr)\quad\text{for all }x\in\mathbb R^n . \notag
  \end{align}
  Let $c_s \coloneqq \|A_\theta s\|$. Note that $c_s$ is finite for any finite $s$. For $\|w\|=R$ we have $\|A_\theta s+w\|\le R+c_s$, hence, one can show that
  \begin{align}
    \|g(A_\theta s+w)\|
    \le C_0\bigl(1+(R+c_s)^2\bigr)
    \le C_1\,(1+R^2) \notag
  \end{align}
  for $C_1:=C_0 (2+c_s^2)$ that is independent of $R$. Therefore,
  \begin{align}
    |B_R| & \le \sup_{\|w\|=R} p_w(w)\;\int_{S_R}\|g(A_\theta s+w)\|  \mathrm{d} S \notag \\
          & \le C_1(1+R^2)\,|S^{n-1}|\,R^{n-1}\,\sup_{\|w\|=R} p_w(w), \notag
  \end{align}
  where $|S^{n-1}|$ is the surface area of the unit sphere in $\mathbb R^n$. Thus, since $1+R^2\le 2R^2$ for $R\ge 1$,
  \begin{align}
    |B_R|\le C_2\, R^{n+1}\,\sup_{\|w\|=R} p_w(w). \label{eq:BR-bound}
  \end{align}
  By Assumption~\ref{assump:vanishing-boundary-flux} (ii),
  \begin{align}
    \lim_{R\to\infty} R^{n+1}\sup_{\|w\|=R} p_w(w)=0,
  \end{align}
  and hence $\lim_{R\to\infty} B_R=0$. This shows that the boundary contribution at infinity vanishes:
  \begin{align}
    \int_{\partial\mathcal W} p_w(w)\, \boldsymbol n(w)\, g(A_\theta s+w) \tran \,\mathrm{d} S
    = \lim_{R\to\infty} B_R = 0 .
  \end{align}
  The claim follows by satisfying either of the two cases. \hfill $\blacksquare$
\end{pf}
\smallskip
\begin{Remark}
  \label{rem:discussion-bounded-i}
  Case (i) (bounded support). If $\mathcal W\subset\mathbb R^n$ has Lipschitz boundary and $p_w=0$ on $\partial\mathcal W$, the boundary term vanishes. When a given density does not vanish on $\partial\mathcal W$ (e.g., the uniform law on $\mathcal W$), the condition can be enforced by a smooth boundary-layer cutoff. Let $d(w)\coloneqq\operatorname{dist}(w,\partial\mathcal W)$ and choose $\eta\in C^\infty([0,\infty))$ with $0\le\eta\le1$, $\eta(0)=0$, and $\eta(t)=1$ for $t\ge1$. For any $\varepsilon>0$ define
  \begin{align*}
       \eta_\varepsilon(w) &\coloneqq \eta\left(\frac{d(w)}{\varepsilon}\right),\qquad p_\varepsilon(w) \coloneqq \frac{\eta_\varepsilon(w)\,p_w(w)}{Z_\varepsilon},\notag \\
    Z_\varepsilon &\coloneqq \int_{\mathcal W}\eta_\varepsilon(w)\,p_w(w)\, \mathrm{d} w. \notag
  \end{align*}
  Then $0\le \eta_\varepsilon\le1$, $\eta_\varepsilon\to1$ pointwise and in $L^1(\mathcal W)$, and $Z_\varepsilon\nearrow1$ by monotone convergence as $\varepsilon\downarrow0$. Since $\eta_\varepsilon=0$ on $\partial\mathcal W$, $p_\varepsilon=0$ on $\partial\mathcal W$. Finally, as $\varepsilon\downarrow0$, $p_\varepsilon\to p_w$ in total variation:
  \begin{align}
    \|p_\varepsilon-p_w\|_{L^1(\mathcal W)} \le & \|(\eta_\varepsilon-1)p_w\|_{L^1(\mathcal W)}  \notag                                       \\
                                                & \qquad + \Big|\,\tfrac{1}{Z_\varepsilon}-1\Big|\,\|\eta_\varepsilon p_w\|_{L^1(\mathcal W)}
    \xrightarrow[\varepsilon\downarrow0]{}0. \notag
  \end{align}
  Thus Assumption~\ref{assump:vanishing-boundary-flux} holds by this approximation approach.

  Case (ii) (unbounded support). The tail condition \eqref{eq:unbounded-support} ensures the boundary contribution at infinity vanishes, since the flux over $S_R$ is $O\big(R^{n+1}\sup_{\|w\|=R}p_w(w)\big)$ by \eqref{eq:BR-bound} 
  It holds whenever, for some $C,c>0$ and all large $\|w\|$,
  \begin{align}
      p_w(w)\le C\,\exp\big(-c\|w\|^\alpha\big)\quad(\alpha>0). \notag
  \end{align}
  Thus the assumption is satisfied by sub-Gaussian laws (e.g., Gaussian) and sub-exponential laws (e.g., Laplace). 
\end{Remark}

\smallskip

\begin{Lemma}[Jacobian of $P_\theta$ w.r.t. $K$]
  \label{lem:jacobian-of-P-symmetric}
  Assume $\theta$ is $\gamma$-stabilizing (Definition~\ref{Def:Stabilization}), then \eqref{eq:discrete-Lyap} has a unique solution $P_\theta$, and its Jacobian w.r.t.\ $\mathrm{vec}(K)$ is
  \begin{align}
    \frac{\partial\,\mathrm{vec}(P_\theta)}{\partial \theta} = T_\theta^{-1} \Big[(S_\theta\tran \otimes I_n) \,K_{mn} + (I_n \otimes  S_\theta\tran)\Big], \label{eq:jacobian-of-P-symmetric}
  \end{align}
  where $\theta = \mathrm{vec}(K)$, $S_\theta \coloneqq R K - \gamma B\tran P_\theta A_{\theta}\in\mathbb{R}^{m\times n}$, $T_\theta \coloneqq I_{n^2}-\gamma(A_{\theta}\tran \otimes A_{\theta}\tran) \in \mathbb{R}^{n^2\times n^2}$.
\end{Lemma}

\begin{pf}
  Starting from the discounted Lyapunov equation \eqref{eq:discrete-Lyap} and taking differentials,
  \begin{align}
    \mathrm{d}P_\theta
    = \mathrm{d}(K\tran R K) + \gamma\,\mathrm{d}(A_{\theta}\tran P_\theta A_{\theta}). \notag
  \end{align}
  Using $\mathrm{d}A_{\theta}=-B\,\mathrm{d}K$ and the product rule,
  \begin{align}
    \mathrm{d}(K\tran R K)                          & = (\mathrm{d}K)\tran R K + K\tran R\,\mathrm{d}K, \notag                                                                                                                                                \\
    \mathrm{d}(A_{\theta}\tran P_\theta A_{\theta}) & = (\mathrm{d}A_{\theta})\tran P_\theta A_{\theta} \! + \! A_{\theta}\tran(\mathrm{d}P_\theta)A_{\theta} \! +\! A_{\theta}\tran P_\theta\,\mathrm{d}A_{\theta}                                    \notag \\
                                                    & = \! -(\mathrm{d}K)\tran \! B\tran P_\theta A_{\theta} \! +\! A_{\theta}\tran(\mathrm{d}P_\theta)A_{\theta} \! - \! A_{\theta}\tran P_\theta B\,\mathrm{d}K . \notag
  \end{align}
  Substituting and collecting the $\mathrm{d}P_\theta$ terms on the left gives the differential Lyapunov equation
  \begin{align}
    \mathrm{d}P_\theta - \gamma\,A_{\theta}\tran(\mathrm{d}P_\theta)A_{\theta}
    = (\mathrm{d}K)\tran S_\theta + S_\theta\tran \mathrm{d}K \notag
  \end{align}
  Vectorizing and using $\mathrm{vec}(X\tran)\!=\!K_{mn}\mathrm{vec}(X)$ identity yields
  \begin{align}
    \big[I_{n^2} - \gamma& (A_{\theta}\tran \otimes A_{\theta}\tran)\big]\mathrm{vec}(\mathrm{d}P_\theta) \notag                                  \\
                 & =  (S_\theta\tran\otimes I_n)K_{mn}\,\mathrm{vec}(\mathrm{d}K) + (I_n\otimes S_\theta\tran)\,\mathrm{vec}(\mathrm{d}K). \notag
  \end{align}
  Since $\theta$ is stabilizing, we have
  $\rho\big(\gamma(A_{\theta}\tran\otimes A_{\theta}\tran)\big)
    =\gamma\,\rho(A_{\theta})^2<1$, hence $T_\theta \coloneq I_{n^2}-\gamma(A_{\theta}\tran\otimes A_{\theta}\tran)$ is invertible. Solving for $\mathrm{vec}(\mathrm{d}P_\theta)$ results in~\eqref{eq:jacobian-of-P-symmetric}.  \hfill $\blacksquare$
\end{pf}

We now prove  Theorem~\ref{thm:Lambda_structure}, which provides a closed-form expression for the exact Hessian for the discounted LQR.

\begin{pf}
  Consider the representation of $ \Lambda(\theta)$ in~\eqref{eq:Lambda-policy-hessian}
  \begin{gather}
    \Lambda(\theta) =\mathbb{E}_{\tau_\theta}\bigl[f_\theta(s)+f_\theta(s)\tran\bigr], \notag \\
    f_\theta(s)  \coloneqq \int_{\mathbb{R}^n} \nabla_{\theta} p\bigl(s' | s,\pi_{\theta}(s)\bigr)\, \nabla_{\theta} V_{\theta}(s')\tran\,\mathrm{d}s'. \label{eq:f-def}
  \end{gather}
  By differentiating \eqref{eq:Kernel} w.r.t.\ $\theta$ we obtain
  \begin{align}
    \nabla_{\theta} p\big(s' | s,\pi_{\theta}(s)\big)
     & = \nabla_{\theta} \,p_{w}\big(s'-A_{\theta}s\big) \notag            \\
     & = \big(-\nabla_{\theta}(A_{\theta}s)\big) \,
    \nabla_{w}p_{w}(w) \big|_{w=s'-A_{\theta}s} \notag                     \\
     & = (s \otimes B\tran) \,\nabla_{w}p_{w}(w) \big|_{w=s'-A_{\theta}s},
    \label{eq:gradTransition}
  \end{align}
  where we used $\nabla_{\theta}(A_{\theta}s) = -\,s\tran \otimes B$ and the Kronecker-product identities. Subsequently, differentiating \eqref{eq:V-theta} yields,
  \begin{align}
    \nabla_{\theta} V_{\theta}(s) & = \nabla_\theta s \tran P_{\theta}s   + \nabla_{\theta} q_{\theta} = g(s) + \nabla_{\theta} q_{\theta}, \label{eq:gradV}
  \end{align}
  where
  \begin{align}
    g(s) & \coloneqq  \nabla_{\theta} s \tran P_{\theta}s =  \big[\, s \tran \tfrac{\partial P_\theta}{\partial \theta_{1}} s \;,\; \ldots \;,\; s \tran \tfrac{\partial P_\theta}{\partial \theta_{mn}} s \,\big]\tran.
    \label{eq:def-g}
  \end{align}
  Substituting \eqref{eq:gradTransition} and \eqref{eq:gradV} into \eqref{eq:f-def} yields
  \begin{align}
    f_\theta(s) & = (s \! \otimes \! B \tran) \int_{\mathbb{R}^n} \! \! \nabla_{w}p_{w}(w) \big|_{w=s'-A_{\theta}s}  \bigl(g(s') \! + \! \nabla_{\theta} q_{\theta}\bigr)\tran \mathrm{d}s' \notag                               \\
                & =(s\! \otimes \! B \tran) \Big(\underbrace{\int_{\mathbb{R}^n}  \! \! \nabla_{w}p_{w}(w) \big|_{w=s'-A_{\theta}s} \, g(s') \tran  \mathrm{d}s'}_{=:I_1} \notag                                                 \\
                & \qquad \qquad + \underbrace{\int_{\mathbb{R}^n} \! \!\nabla_{w}p_{w}(w) \big|_{w=s'-A_{\theta}s} \nabla_{\theta} q_{\theta}\tran \mathrm{d}s'}_{=:I_2}\Big) \raisetag{1.3\baselineskip} \label{eq:f-semifinal}
  \end{align}
  In the following, we focus on evaluating the integrals $I_1$ and $I_2$. The second integral can be simplified as follows
  \begin{align}
    I_2=\int_{\mathbb{R}^n} & \nabla_{w}p_{w}(w)
    \Bigl|_{w = s' - A_{\theta}s}\;
    \nabla_{\theta} q_{\theta}\tran\,\mathrm{d}s'  \notag                       \\[4pt]
                            & =\Bigl(\int_{\mathbb{R}^n} \nabla_{w}p_{w}(w)
    \Bigl|_{w = s' - A_{\theta}s}\,\mathrm{d}s'\Bigr)
    \nabla_{\theta} q_{\theta}\tran       \notag                                \\[4pt]
                            & \stackrel{w = s' - A_{\theta}s}{=}
    \Bigl(\int_{\mathcal{W}}\nabla_{w}p_{w}(w)\,\mathrm{d}w\Bigr)
    \nabla_{\theta} q_{\theta}\tran      \notag                                 \\[4pt]
                            & =\int_{\partial\mathcal W} p_w(w)\,\boldsymbol n(w)\,\mathrm dS \;
    \nabla_{\theta} q_{\theta}\tran=0.
    \label{eq:constTermZero_general}
  \end{align}
The final equality follows from the boundary-vanishing property of $p_{w}$ on $\partial\mathcal{W}$ (see Assumption~\ref{assump:vanishing-boundary-flux}). Note that $q_{\theta}$ depends on $\theta$ only through $P_{\theta}$, and $P_{\theta}$ is the unique solution of the discounted discrete-time Lyapunov equation. Therefore map $\theta \mapsto P_\theta$ is smooth for $\gamma$-stabilizing $\theta$ and it yields that $q_{\theta}$ is also smooth with bounded $\nabla_{\theta} q_{\theta}$.

  To evaluate $I_1$, we invoke Lemma~\ref{thm:vanishing-boundary-flux}. Applying integration by parts in the variable $w$ yields
  \begin{align}
    I_1 =                              & \int_{\mathbb{R}^n}\nabla_{w}  p_{w}(w)\big|_{w=s'-A_{\theta}s}\, g(s')\tran \,\mathrm{d}s' \notag                         \\
    \stackrel{w = s' - A_{\theta}s}{=} & \int_{\mathcal W} \nabla_{w} p_{w}(w)\; g\left(A_{\theta}s+w\right)\tran \,\mathrm{d}w \notag                              \\[4pt]
    =                                  & \int_{\partial\mathcal W} p_w(w)\, \boldsymbol{n}(w) g( A_{\theta} s + w)\tran \,  \mathrm{d} S  \notag                                  \\
                                       & \quad -  \int_{\mathcal W} p_{w}(w)\; \big(\nabla_{w} g\left(A_{\theta}s+w\right)\big)\tran \, \mathrm{d}w, \label{eq:integrationByParts}
  \end{align}
  By Theorem~\ref{thm:vanishing-boundary-flux}, the first term on the right-hand side of \eqref{eq:integrationByParts} vanishes.
  To evaluate the second term, we write
  \begin{align}
    \big(\nabla_w g(A_\theta s+w)\big)_i = (A_\theta s+w)\tran\!\left(\frac{\partial P_\theta}{\partial \theta_i}  +\left(\frac{\partial P_\theta}{\partial \theta_i}\right)\tran\right),  \notag
  \end{align}
  Since $\dfrac{\partial P_\theta}{\partial \theta_i}$ does not depend on $w$ and $\mathbb{E}[w]=0$,
  \begin{align}
    \int p_w(w)\,(A_\theta s+w)\tran\,\mathrm{d}w
    = s\tran A_\theta\tran. \notag
  \end{align}
  Hence
  \begin{align}
    \mathbb{E}_w\!\left[\big(\nabla_w g(A_\theta s+w)\big)_i\right] = s\tran A_\theta\tran \!\left( \frac{\partial P_\theta}{\partial \theta_i}+ \left(\frac{\partial P_\theta}{\partial \theta_i}\right)\tran \right). \notag
  \end{align}

  Define, for $i=1,\dots,mn$, the vectors $\lambda_{\theta_i}\in\mathbb{R}^n$ by
  \begin{align}
    \lambda_{\theta_i} \coloneq \left( \dfrac{\partial P_\theta}{\partial \theta_i}+ \left(\dfrac{\partial P_\theta}{\partial \theta_i}\right)\tran \right) A_\theta s. \notag
  \end{align}

  Then we can write the gradient of the integral in \eqref{eq:integrationByParts} as
  \begin{align}
    \int_{\mathcal{W}} p_w(w)\,\nabla_w g(A_\theta s+w)\tran\,\mathrm{d}w = \begin{bmatrix}
                                                                              \lambda_{\theta_1} \; \cdots \; \lambda_{\theta_{mn}}
                                                                            \end{bmatrix}. \label{eq:integrationByParts3-final}
  \end{align}

  Substituting \eqref{eq:integrationByParts3-final} into \eqref{eq:f-semifinal} yields
  \begin{align}
    f_\theta(s)
     & = (s \otimes B\tran)\Bigl[-\int p_{w}(w)\,\nabla_{w} g  \bigl(A_{\theta}s+w\bigr)\tran \,\mathrm{d}w\Bigr] \notag \\
     & = -\,(s\otimes B\tran)
    \begin{bmatrix}
      \lambda_{\theta_{1}} & \cdots & \lambda_{\theta_{mn}}
    \end{bmatrix} \notag                                       \\
     & = -\,
    \begin{bmatrix}
      \mathrm{vec} \bigl(B\tran \lambda_{\theta_{1}} s\tran\bigr)
       & \cdots &
      \mathrm{vec} \bigl(B\tran \lambda_{\theta_{mn}} s\tran\bigr)
    \end{bmatrix}, \notag
  \end{align}
  where, in the last equality, the vec--Kronecker identity is applied columnwise. For subsequent use, define the constant matrices $M_i \in \mathbb{R}^{m\times n} $ by
  \begin{align}
    M_i \coloneqq B\tran \Bigl(\frac{\partial P_\theta}{\partial \theta_{i}}
    + \frac{\partial P_\theta}{\partial \theta_{i}}\tran \Bigr) A_{\theta}
    ,
    \; i=1,\dots,mn.
    \label{eq:def-Mi}
  \end{align}

  Then we have $B\tran \lambda_{\theta_i} = M_i s$, and
  \begin{align}
    f_\theta(s)
     & = -\,
    \begin{bmatrix}
      \mathrm{vec}\bigl(M_1 s s\tran \bigr) \;\cdots\;
      \mathrm{vec}\bigl(M_{mn} s s\tran \bigr)
    \end{bmatrix} \notag \\
     & = -\,
    \begin{bmatrix}
      (I_n \otimes M_1)\,\mathrm{vec}(s s\tran ) \cdots
      (I_n \otimes M_{mn})\,\mathrm{vec}(s s\tran )
    \end{bmatrix}.
    \notag
  \end{align}
  With $\Sigma_\theta  \coloneqq  \mathbb{E}_{\tau_\theta}[s s\tran ]$ and the linearity of expectation,
  \begin{align}
    \mathbb{E}_{\tau_\theta}[f_\theta(s)]
     & = -
    \begin{bmatrix}
      (I_n \!\otimes\! M_1)\mathrm{vec}(\Sigma_\theta) \cdots
      (I_n \!\otimes\! M_{mn})\mathrm{vec}(\Sigma_\theta)
    \end{bmatrix} \notag \\
     & = -
    \begin{bmatrix}
      \mathrm{vec}(M_1 \Sigma_\theta) \cdots \mathrm{vec}(M_{mn} \Sigma_\theta)
    \end{bmatrix}
    \notag                                                 \\
     & = -
    \begin{bmatrix}
      (\Sigma_\theta \!\otimes\! I_m)\mathrm{vec}(M_1) \cdots
      (\Sigma_\theta \!\otimes\! I_m)\mathrm{vec}(M_{mn})
    \end{bmatrix} \notag \\
     & = -(\Sigma_\theta \!\otimes\! I_m)
    \begin{bmatrix}
      \mathrm{vec}(M_1) \cdots \mathrm{vec}(M_{mn})
    \end{bmatrix} . \label{eq:MiSig}
  \end{align}
  In the second equality of \eqref{eq:MiSig} we used $\mathrm{vec}(M\Sigma)=(I\otimes M)\mathrm{vec}(\Sigma)$, and in the third we used, $\mathrm{vec}(I_m M_i \Sigma_\theta) = (\Sigma_\theta\tran  \otimes I_m)\mathrm{vec}(M_i)$, and the symmetry of $\Sigma_\theta\tran =\Sigma_\theta$.
  Note that $\mathrm{vec}(M_i) = (A_\theta\tran \otimes B\tran)\Big(\mathrm{vec}\Big(\frac{\partial P_\theta}{\partial \theta_i}
+ \big(\frac{\partial P_\theta}{\partial \theta_i}\big)\tran\Big)\Big)$ 
  by \eqref{eq:def-Mi} and the vec--Kronecker identity. Thus,
  \begin{align}
   & \mathbb{E}_{\tau_\theta}[f_\theta(s)]\!= \!-(\Sigma_\theta \otimes I_m)(A_\theta\tran \otimes B\tran) (I_{n^2} + K_{nn})\frac{\partial\,\mathrm{vec}(P_\theta)}{\partial\,\theta}   \notag \\
                                          & = -(\Sigma_\theta A_\theta\tran \otimes B\tran) (I_{n^2} + K_{nn}) \frac{\partial\,\mathrm{vec}(P_\theta)}{\partial\,\theta},\qquad  \raisetag{1\baselineskip} \label{eq:f-final-app} 
  \end{align}
  where $   \frac{\partial\,\mathrm{vec}(P_\theta)}{\partial\,\theta}=\begin{bmatrix}
                                                                  \mathrm{vec}\big(\tfrac{\partial P_\theta}{\partial \theta_1}\big) & \cdots & \mathrm{vec}\big(\tfrac{\partial P_\theta}{\partial \theta_{mn}}\big)
                      \end{bmatrix}$, and in the last equality we used the mixed-product property of Kronecker products.
  Since $P_\theta$ is symmetric, each column of its Jacobian $\tfrac{\partial\,\mathrm{vec}(P_\theta)}{\partial\theta}$ is the vectorization of a symmetric matrix. Consequently, these columns are fixed by the commutation matrix $K_{nn}$:
  \begin{align}
    K_{nn}\,\frac{\partial\,\mathrm{vec}(P_\theta)}{\partial\theta}
    = \frac{\partial\,\mathrm{vec}(P\tran_\theta)}{\partial\theta}= \frac{\partial\,\mathrm{vec}(P_\theta)}{\partial\theta}=I_{n^2}\frac{\partial\,\mathrm{vec}(P_\theta)}{\partial\theta}. \notag
  \end{align}
  Therefore, \eqref{eq:f-final-app} reads as 
  \begin{align*}
    \mathbb{E}_{\tau_\theta}[f_\theta(s)]&=-(\Sigma_\theta A_\theta\tran \otimes B\tran) (I_{n^2} + I_{n^2} ) \frac{\partial\,\mathrm{vec}(P_\theta)}{\partial\,\theta}\\
     & = -2\,(\Sigma_\theta A_\theta\tran \otimes B\tran) \frac{\partial\,\mathrm{vec}(P_\theta)}{\partial\theta}. 
  \end{align*}
  Combining the above with its transpose gives~\eqref{eq:Lambda-final}. \hfill $\blacksquare$
\end{pf}

\end{document}